\documentclass[copyright,creativecommons,noderivs,noncommercial]{eptcs}
\providecommand{\event}{EXPRESS 2009} 
\providecommand{\volume}{??}
\providecommand{\anno}{2009}
\providecommand{\firstpage}{1}
\providecommand{\eid}{??}

\usepackage[latin1]{inputenc}
\usepackage{amssymb}
\usepackage{amsmath}
\usepackage{graphics}
\usepackage{breakurl}

\newcommand\ff{f\!\!f}
\newcommand\tr{t\!t}

\newcommand{\Po}{{\cal P}}

\newcommand{\Lo}{{\cal L}}

\newcommand{\ioi}{\Leftrightarrow}

\newcommand{\der}[1]{\ensuremath{\stackrel{#1}{\longrightarrow}}}
\newcommand{\ind}[1]{\ensuremath{\stackrel{#1}{\cdots}}}
\newcommand{\divg}{\uparrow}
\newcommand{\conv}{\downarrow}
\newcommand{\fun}{\rightarrow}

\newcommand{\sqlt}{\sqsubseteq}
\newcommand{\sqgt}{\sqsupseteq}

\newcommand{\Psys}{P}
\newcommand{\F}{{\cal F}}
\newcommand{\Proc}{{\bf P}}
\newcommand{\Act}{{\bf A}}
\newcommand{\Id}{\mathcal{I}}

\newcommand{\Var}{\mathit{Var}}
\newcommand{\Env}{\mathit{Env}}

\DeclareMathOperator{\fix}{\mathit{fix}}
\DeclareMathOperator{\Fix}{\mathit{Fix}}

\newcommand\lb {[\![}
\newcommand\rb{]\!]}

\newcommand{\putaway}[1]{}
\newcommand{\must}[1]{[ #1 ]}
\newcommand{\may}[1]{\langle #1 \rangle}

\newcommand{\sem}[1]{\relax\ifmmode \lb #1 \rb \else $\lb #1 \rb$ \fi}
\newcommand{\den}[1]{\sem{ #1 }}

\newcommand{\qed}{\nobreak \ifvmode \relax \else
      \ifdim\lastskip<1.5em \hskip-\lastskip
      \hskip1.5em plus0em minus0.5em \fi \nobreak
      \vrule height0.75em width0.5em depth0.25em\fi}
      
\newcommand{\casedef}[3]
{\left\{
\begin{array}{lr}
#1 &\mbox{ if } #2\\
#3 &\mbox{ otherwise. }
\end{array}\right.}

\newtheorem{theorem}{Theorem}[section]
\newtheorem{lemma}[theorem]{Lemma}
\newtheorem{proposition}[theorem]{Proposition}
\newtheorem{corollary}[theorem]{Corollary}
\newtheorem{definition}[theorem]{Definition}

\newenvironment{proof}[1][Proof]{\begin{trivlist}
\item[\hskip \labelsep {\bfseries #1}]}{\end{trivlist}}

\title{Characteristic Formulae for Fixed-Point Semantics: \\ A
 General Framework\thanks{The work of the authors has been partially
 supported by the project ``New Developments in Operational
 Semantics'' (nr.~080039021) of the Icelandic Research Fund. Joshua
 Sack has been further supported by a grant from Reykjavik
 University's Development Fund.}} 

\author{Luca Aceto \quad
Anna Ingolfsdottir \quad 
Joshua Sack
\institute{School of Computer Science, Reykjavik University,
\\ Kringlan 1, IS-103 Reykjavik, Iceland} 
\email{luca@ru.is, annai@ru.is, joshua.sack@gmail.com}
}

\def\authorrunning{Aceto, Ingolfsdottir and Sack}
\def\titlerunning{Characteristic Formulae for Fixed-Point Semantics: A
 General Framework}

\begin{document}
\maketitle

\begin{abstract}
The literature on concurrency theory offers a wealth of examples of
characteristic-formula constructions for various behavioural relations
over finite labelled transition systems and Kripke structures that are
defined in terms of fixed points of suitable functions. Such
constructions and their proofs of correctness have been developed
independently, but have a common underlying structure. This study
provides a general view of characteristic formulae that are expressed
in terms of logics with a facility for the recursive definition of
formulae. It is shown how several examples of characteristic-formula
constructions from the literature can be recovered as instances of the
proposed general framework, and how the framework can be used to yield
novel constructions.
\end{abstract}

\section{Introduction}

Various types of automata are fundamental formalisms for the
description of the behaviour of computing systems.  For instance, a
widely used model of computation is that of {\em labelled transition
systems} (LTSs)~\cite{Ke76}.  LTSs underlie Plotkin's Structural
Operational Semantics~\cite{Plotkin04a} and, following Milner's
pioneering work on CCS~\cite{Mi89}, are by now the formalism of choice
for describing the semantics of various process description languages.

Since automata like LTSs can be used for describing specifications of
process behaviours as well as their implementations, an important
ingredient in their theory is a notion of behavioural equivalence or
preorder between (states of) LTSs. A behavioural equivalence describes
formally when (states of) LTSs afford the same `observations', in
some appropriate technical sense. On the other hand, a behavioural
preorder is a possible formal embodiment of the idea that (a state in)
an LTS affords at least as many `observations' as another
one. Taking the classic point of view that an implementation correctly
implements a specification when each of its observations is allowed by
the specification, behavioural preorders may therefore be used to
establish the correctness of implementations with respect to their
specifications, and to support the stepwise refinement of
specifications into implementations. 

The lack of consensus on what constitutes an appropriate notion of
observable behaviour for reactive systems has led to a large number of
proposals for behavioural equivalences and preorders for concurrent
processes.  In his by now classic paper~\cite{vG2001},
van Glabbeek presented a taxonomy of extant behavioural preorders and
equivalences for processes.

The approach to the specification and verification of reactive systems
in which automata like LTSs are used to describe both implementations
and specifications of reactive systems is often referred to as {\em
implementation verification} or {\em equivalence checking}.

An alternative approach to the specification and verification of
reactive systems is that of {\em model
checking}~\cite{ClEm81,ClarkeGP1999,QS81}. In this approach, automata
are still the formalism of choice for the description of the actual
behaviour of a concurrent system. However, specifications of the
expected behaviour of a system are now expressed using a suitable
logic, for instance, a modal or temporal
logic~\cite{Emerson1990,Pnueli77}. Verifying whether a concurrent
process conforms to its specification expressed as a formula in the
logic amounts to checking whether the automaton describing the
behaviour of the process is a model of the formula.

It is natural to wonder what the connection between these two
approaches to the specification and verification of concurrent
computation is. A classic, and most satisfying, result in the theory
of concurrency is the characterization theorem of bisimulation
equivalence~\cite{Mi89,Pa81} in terms of Hennessy-Milner logic (HML)
due to Hennessy and Milner~\cite{HM85}. This theorem states that two
bisimilar processes satisfy the same formulae in Hennessy-Milner
logic, and if the processes satisfy a technical finiteness condition,
then they are also bisimilar when they satisfy the same formulae in
the logic. This means that, for bisimilarity and HML, the process
equivalence induced by the logic coincides with behavioural
equivalence, and that, whenever two processes are {\em not}
equivalent, we can always find a formula in HML that witnesses a
reason why they are not. This distinguishing formula is useful for
debugging purposes, and can be algorithmically constructed for finite
processes---see, e.g., \cite{Kor91,MaSt93}. 

The characterization theorem of Hennessy and Milner is, however, less
useful if we are interested in using it directly to establish when two
processes are behaviourally equivalent using model checking. Indeed,
that theorem seems to indicate that to show that two processes are
equivalent we need to check that they satisfy the same formulae
expressible in the logic, and there are countably many such formulae,
even modulo logical equivalence. Is it possible to find a {\em single}
formula that characterizes the bisimulation equivalence class of a
process $p$---in the sense that any process is bisimilar to $p$ if,
and only if, it affords that property? Such a formula, if it exists,
is called a {\em characteristic formula}. When a characteristic
formula for a process modulo a given notion of behavioural equivalence
or preorder can be algorithmically constructed, implementation
verification can be reduced to model checking, and we can translate
automata to logic. (An investigation of the model checking problems
that can be reduced to implementation verification may, for instance,
be found in the paper~\cite{BoudolL1992}.)

Characteristic formulae provide a very elegant connection between
automata and logic, and between implementation verification and model
checking. But, can they be constructed for natural, and suitably
expressive, automata-based models and known logics of computation?  To
the best of our knowledge, this natural question was first addressed
in the literature on concurrency theory in the paper~\cite{GrafS86a}.
In that paper, Graf and Sifakis offered a translation from
recursion-free terms of Milner's CCS~\cite{Mi89} into formulae of a
modal language representing their equivalence class with respect to
observational congruence.

Can one characterize the equivalence class of an arbitrary finite
process---for instance one described in the regular fragment of
CCS---up to bisimilarity using HML? The answer is negative because
each formula in that logic can only describe a finite fragment of the
initial behaviour of a process---see, for instance,~\cite{AILS2007}
for a textbook presentation.  However, as shown in,
e.g.,~\cite{IGZ,SteffenI94}, adding a facility for the recursive
definition of formulae to (variants of) HML yields a logic that is
powerful enough to support the construction of characteristic formulae
for various types of finite processes modulo notions of behavioural
equivalence or preorder.

Following on the work presented in those original references, the
literature on concurrency theory offers by now a wealth of examples of
characteristic-formula constructions for various behavioural relations
over finite labelled transition systems, Kripke structures and timed
automata that are defined in terms of fixed points of suitable
functions. (See, for instance, the
references~\cite{AcetoIPP00,BCG88,CleavelandS1991,FecherS:bib:05:CFUTS,llw95,LS92,MullerOlm1998}.)
Such constructions and their proofs of correctness have been developed
independently, but have a common underlying structure. It is therefore
natural to ask oneself whether one can provide a general framework
within which some of the aforementioned results can be recovered from
general principles that isolate the common properties that lie at the
heart of all the specific constructions presented in the
literature. Not only do such general principles allow us to recover
extant constructions in a principled fashion, but they may also yield
novel characteristic-formula constructions `for free'.

In this study, we offer a general view of characteristic formulae that
are expressed in terms of logics with a facility for the recursive
definition of formulae. The proposed framework applies to behavioural
relations that are defined as fixed points of suitable monotonic
functions. Examples of such relations are those belonging to the
family of bisimulation- and simulation-based semantics. We show that
if, in a suitable technical sense defined in
Section~\ref{section:theorem}, a recursively defined logical formula expresses
the function underlying the definition of a behavioural relation,
then the largest interpretation of that formula is exactly the
characteristic formula for the derived behavioural relation. (See
Theorem~\ref{charform}.)  Using this result, we are able to recover,
as instances of the proposed general framework and essentially for
free, several examples of characteristic-formula constructions from
the literature. In particular, we focus on simulation~\cite{Pa81},
bisimulation~\cite{Mi89,Pa81}, ready
simulation~\cite{BloomIM1995,LarsenS91}, and prebisimulation
semantics~\cite{Mi81}. In addition, we show that the framework can be
used to yield novel constructions. By way of example, we provide
characteristic formulae for back and forth bisimilarity, back and
forth bisimilarity with indistinguishable states~\cite{MousaviLPAR07}
and extended simulation semantics~\cite{Thom87}.

We trust that the general view of characteristic-formula constructions
we provide in this article will offer a framework for the derivation
of many more such results and for explaining the reasons underlying
the success of extant constructions of this kind in the literature.
\paragraph{Roadmap of the paper} 
The paper is organized as follows. In Section \ref{background} we
describe the theoretical background the paper relies on. Section
\ref{section:theorem} contains the main theorem of the paper whereas
Section \ref{applications} is devoted to its applications. Finally in
Section \ref{conclusion} we give some concluding remarks.

\section{Some theoretical  background}\label{background}
In this section we provide the theoretical background needed in the
paper.

\subsection{Fixed points, prefixed points, and postfixed points}

Let $\mathbf{X}$ be a set and $\F:\Po(\mathbf{X}) \fun\Po(\mathbf{X})$
be a monotonic function, i.e. a function that preserves the order
$\subseteq$.  A set $S\subseteq \mathbf{X}$ is a \emph{fixed point} of
$\F$ if $\F(S) = S$, a \emph{prefixed point} of $\F$ if
$\F(S)\subseteq S$, and a \emph{postfixed point} of $\F$ if
$S\subseteq \F(S)$.  By Tarski's fixed-point theorem \cite{Ta55}, $\F$
has both a unique largest fixed point given by union of all its postfixed
points:
\[
\Fix\F=\bigcup\{S\subseteq \mathbf{X} \mid S\subseteq\F(S)\}\mbox{}
\]
and a unique least one given by the intersection of all of its prefixed points:
\[
\fix\F=\bigcap\{S\subseteq \mathbf{X} \mid \F(S)\subseteq S \}.\mbox{}
\]
The largest fixed points of such functions are commonly used as the
basis for many co-inductively defined behavioural semantics as we will
see later in the paper. 

\subsection{Logic}
We assume a formal specification language or logic $\Lo(\Var)$
defined over a set of variables $\Var$ and ranged over by $F$, possibly
with subscripts.   We often write
$\Lo$ for $\Lo(\Var)$ if the set of variables is clear from the context.

We also involve a function $D:\Var \fun\Lo$ called a
\emph{declaration}; the declaration can be viewed as providing us with
a system of equations over $\Var$ that decides the meaning of each
variable.

We interpret the language over a given set $\Proc$. For this purpose we
assign to each variable a set of elements in $\Proc$ for which this
variable holds true. To cater for this, we use a function $\sigma:
\Var \fun \Po(\Proc)$, called an {\em environment}.  Let $\Env(\Var)$
be the set of all environments (ranged over by $\sigma$) whose domain
is $\Var$. Again, we often write $\Env$ for $\Env(\Var)$ if the
set of variables is clear from the context.  Ordered under the pointwise set
inclusion, which we again refer to as $\subseteq$, $\Env$ is a
complete lattice with respect to the induced pointwise $\bigcup$ and
$\bigcap$ as the least upper bound and the greatest lower bound
respectively.  

For each environment $\sigma:\Var \to \Po(\Proc)$, we let $\models\,\,
\subseteq (\Env\times \Proc) \times \Lo$ be a relation, with the
condition that for each $X\in \Var$, we have $(\sigma, p) \models X$
if and only if $p\in \sigma(X)$.  A language is monotonic if for every
$p\in \Proc$, formula $F$ and environments $\sigma_1$ and $\sigma_2$,
$(\sigma_1, p) \models F$ implies $(\sigma_2, p) \models F$, whenever
$\sigma_1\subseteq \sigma_2$.

A declaration function
$D$ induces a function $\den{D}:\Env\fun \Env$ defined by
$$(\den{D}\sigma)(X)= \{p\mid (\sigma,p) \models D(X)\}.$$
Monotonicity of the language guarantees that $\den{D}$ is monotonic.
We say that $D$ is monotonic whenever $\den{D}$ is.  If $\den{D}$ is
monotonic it has both a largest and a least fixed point over $Env$
which we refer to as $\sigma^{D}_{\max}$ (the largest interpretation
of $D$) and $\sigma^D_{\min}$ (the least interpretation of $D$)
respectively.  We also drop the superscript $D$ as the meaning should
be clear from the context.

In this study we assume that $\Var$ is indexed over some set $\Proc$,
i.~e.~$\Var=\{X_q \mid q\in\Proc\}$. We say that the largest
interpretation of a declaration $D$ gives the characteristic formula
for a binary relation $S$ over $\Proc$ if, for all $p,q\in\Proc$,
\[
p\in\sigma^{D}_{\max}(X_q) \ioi (p,q)\in S .
\]
Characteristic formulae given in terms of least interpretations are
defined similarly.

\section{Expressiveness up-to a relation and characteristic formula}
\label{section:theorem}
In this section we show how we can derive a characteristic formula for
 a semantic relation directly from the logical description of the
 monotonic function that defines the relation. To obtain this, for
 $S\subseteq\Proc\times\Proc$, we define the environment $\sigma_S$ by
\[
\sigma_S(X_q)=\{p\in\Proc \mid (p,q)\in S\},
\]
for each $q\in \Proc$.
\begin{definition}
Given a function $\F:\Po(\Proc\times \Proc)\to \Po(\Proc \times
\Proc)$, a declaration $D:\Var\fun\Lo$ and a set $S\subseteq
\Proc\times \Proc$, we say that \emph{$D$ expresses $\F$ up to $S$} if
for any $p,q\in \Proc$,
\begin{displaymath}
(\sigma_S, p) \models D(X_q) \Leftrightarrow\,\,(p,q)\in\F(S).
\end{displaymath}
\end{definition}
Next we prove that when $D$ expresses $\F$ up to $S$ then the post-
and prefixed points of $\sem{D}$ correspond to the post- and
prefixed points for $\F$.
\begin{lemma}\label{lemma:charform}
Assume that $\F:\Po(\Proc\times \Proc)\to \Po(\Proc \times \Proc)$ is
a monotonic function and $D:\Var\fun\Lo$ is a monotonic declaration
such that $D$ expresses $\F$ up to $S$.  Then  the following hold:
\begin{equation} \label{reference1}
S\subseteq\F(S)\,\ioi\,\sigma_S\subseteq\den{D}\sigma_S,
\end{equation}
and
\begin{equation}\label{reference2}
\F(S)\subseteq S\,\ioi\,\den{D}\sigma_S\subseteq\sigma_S.
\end{equation}
\end{lemma}
\begin{proof}
Assume that $\F:\Po(\Proc\times \Proc)\to \Po(\Proc \times \Proc)$ is
a monotonic function, $D$ is a monotonic declaration, and $S$ is a
relation for which $D$ expresses $\F$ up to $S$.  We will only prove
(\ref{reference1}) as the argument for (\ref{reference2}) is similar.
Towards proving (\ref{reference1}), first assume that
$S\subseteq\F(S)$ and that $u\in \sigma_S(X_v)$ for some $u,v\in
\Proc$. By the definition of $\sigma_S$, $(u,v)\in
S\subseteq\F(S)$. Then,  as $D$ expresses $\F$ up to $S$,
$u\in\den{D}\sigma_S(X_v)$.

Next assume that $\sigma_S\subseteq\den{D}\sigma_S$ and that $(u,v)\in
S$. This implies that
$u\in\sigma_S(X_v)\subseteq\den{D}\sigma_S(X_v)$. As $D$ expresses
$\F$ up to $S$, we have that $(u,v)\in\F(S)$ as desired.  \hfill
$\Box$
\end{proof}
The following theorem states that if the declaration $D$ expresses
$\F$ up to $S$ for every relation $S$ then the largest fixed point of
$\den{D}$ characterizes the largest fixed point of $\F$. This means
that $D$, under the largest interpretation, defines the characteristic
formula for $\Fix\F$. For the sake of completeness, we prove a similar
result for the least fixed points although at this point we have not
found any concrete applications of that result.

\begin{theorem}\label{charform}
Assume that $\F:\Po(\Proc\times \Proc)\to \Po(\Proc \times \Proc)$ is
a monotonic function and $D:\Var\fun\Lo$ is a monotonic declaration
such that $D$ expresses $\F$ up to $S$ for all $S\subseteq \Proc\times
\Proc$.  Then for all $p,q\in\Proc$,
\begin{enumerate}
\item 
$
(\sigma_{\max}, p) \models X_q\Leftrightarrow\,\,(p,q)\in \Fix\F,
$
and
\item 
$
(\sigma_{\min}, p) \models X_q\Leftrightarrow\,\,(p,q)\in \fix\F.
$
\end{enumerate}
\end{theorem}
\begin{proof}
Assume that $\F:\Po(\Proc\times \Proc)\to \Po(\Proc \times \Proc)$ is
a monotonic function and $D$ is a monotonic declaration that
expresses $\F$ up to $S$ for every relation $S$;
that is, for each $S\subseteq\Proc\times \Proc$ and $p,q\in\Proc$, 
\[
(\sigma_S, p) \models D(X_q) \Leftrightarrow\,\,(p,q)\in\F(S).
\]
\begin{enumerate}
\item
We prove that for each $p,q\in\Proc$,
\[
(\sigma_{\max} ,p) \models X_q\Leftrightarrow\,\,(p,q)\in \Fix\F,
\]
or equivalently that
\[
p\in\sigma_{\max}(X_q)\Leftrightarrow\,\,(p,q)\in \Fix\F.
\]
We prove each of the implications separately.
\begin{description}
\item{$\Rightarrow$:} First define $T=\{(u,v)\mid
u\in\sigma_{\max}(X_v)\}$. Then for all $u,v\in\Proc$,
\[
u\in\sigma_{T}(X_v)\ioi (u,v)\in T\ioi u\in\sigma_{\max}(X_v).
\]
This implies that $\sigma_{\max}=\sigma_T$; in particular $\sigma_T$ is a
fixed point, and hence a postfixed point of $\den{D}$.

Towards proving the statement, assume that $p\in\sigma_{\max}(X_q)$.
Then as $\sigma_T$ is a postfixed point of $\den{D}$ and
$\sigma_{\max} = \sigma_T$,
\[
p\in\sigma_T(X_q)\subseteq\den{D}\sigma_T(X_q).
\]
Since $D$ expresses $\F$ up to $T$, we may apply (\ref{reference1}) in
Lemma \ref{lemma:charform} to obtain $T\subseteq \F(T)$.  Therefore,
as $(p,q)\in T$, we have that $(p,q)\in \Fix \F$.

\item{$\Leftarrow:$} Assume $(p,q)\in \Fix\F$. As $\Fix\F=\F(\Fix\F)$
and $D$ expresses $\F$ up to $\Fix\F$, by (\ref{reference1}) in Lemma
\ref{lemma:charform}, we have that
\[
\sigma_{\Fix\F}(X_q)\subseteq\den{D}\sigma_{\Fix\F}(X_q).
\]
As $(p,q)\in \Fix\F$ implies $p\in \sigma_{\Fix\F}(X_q)$, this in turn,
implies that $p\in\sigma_{\max}(X_q)$.
\end{description}
 \item Now we prove that for each $p,q\in\Proc$,
\[
(\sigma_{\min},p) \models X_q\Leftrightarrow\,\,(p,q)\in \fix\F,
\]
or equivalently that
\[
p \in\sigma_{\min} (X_q)\Leftrightarrow\,\,(p,q)\in \fix\F.
\]
We prove each of the implications separately.
\begin{description}
\item{$\Rightarrow$: } Assume
$$p\in\sigma_{\min}(X_q)=\bigcap_{\den{D}\sigma\subseteq\sigma}\sigma(X_q).$$
In other words, for all $\sigma$, $\den{D}\sigma\subseteq\sigma$
implies that $p\in\sigma(X_q)$. We will prove that $(p,q)\in
\fix\F=\bigcap_{\F(S)\subseteq S}S$ or equivalently that for all $S$,
$\F(S)\subseteq S$ implies that $(p,q)\in S$. Towards proving this,
assume that $\F(S)\subseteq S$.  We aim at showing that $(p,q)\in S$. 
Since  $D$
expresses $\F$ up to $S$, 
by (\ref{reference2}) in Lemma \ref{lemma:charform},
$\den{D}\sigma_S\subseteq\sigma_S$. By the assumption above, this
implies that $p\in\sigma_S(X_q)$, or equivalently $(p,q)\in S$.
\item{$\Leftarrow$: } Assume that $(p,q)\in
\fix\F=\bigcap_{\F(S)\subseteq S}S$, or equivalently that for all
$S\subseteq\Proc$, $\F(S)\subseteq S$ implies that $(p,q)\in S$. We
will prove that
$$p\in\sigma_{\min}(X_q)=\bigcap_{\den{D}\sigma\subseteq\sigma}\sigma(X_q).$$
To prove this, it is sufficient to prove that for each environment $\sigma$,
\[
\den{D}\sigma\subseteq\sigma\mbox{ implies }p\in\sigma(X_q).
\]
Towards proving this, assume that
$\den{D}\sigma\subseteq\sigma$. Define
$$T=\{(u,v) \mid u\in\sigma(X_v)\}.$$ Then $\sigma_T=\sigma$ and
therefore $\den{D}\sigma_T\subseteq\sigma_T$. 
Since  $D$
expresses $\F$ up to $T$, by (\ref{reference2}) in
Lemma \ref{lemma:charform}, this implies that $\F(T)\subseteq T$. This
in turn implies that $(p,q)\in T$ and therefore that
$p\in\sigma_T(X_q)=\sigma(X_q)$ as we wanted to prove. \hfill $\Box$
\end{description}
\end{enumerate}
\end{proof}
We have the following corollary.
\begin{corollary}\label{cor:charform}
If a declaration $D$ expresses a monotonic function $\F$ (over
$\Po(\Proc\times\Proc)$) up to any relation, then the largest
interpretation of $D$ gives the characteristic formula for $\Fix \F$.
\end{corollary}
\subsection{Some Observations about Fixed Points}
As usual, for a relation $S\subseteq\Proc\times\Proc$, we let
$S^{-1}=\{(p,q) \mid (q,p)\in S\}$. Furthermore we define 
$\F^*(S)=(\F(S^{-1}))^{-1}$. We have the following properties:
\begin{lemma}\label{inverse}
The following hold.
\begin{enumerate}
\item The function $S\mapsto S^{-1}$ is monotonic and bijective.
\item If $\F$ is monotonic then $\F^{*}$ is monotonic.
\item $\Fix \F^*=(\Fix \F)^{-1}$.
\end{enumerate}
\end{lemma}
\begin{proof}
The proofs of $1-2$ follow directly from the definition of
$S^{-1}$ and $\F^*$ and we only give the details of the proof of $3$.
We proceed with this proof as follows:
\[
\begin{array}{l}
(p,q)\in \Fix \F^*\ioi\exists S.(p,q)\in S\subseteq\F^*(S)\ioi\\\\
\exists S. (p,q)\in S\subseteq(\F(S^{-1}))^{-1}\ioi\\\\
\exists S. (q,p)\in S^{-1}\subseteq\F(S^{-1})\ioi\\\\
\exists R. (q,p)\in R\subseteq\F(R)\ioi \\\\
(q,p)\in \Fix \F\ioi (p,q)\in (\Fix\F)^{-1}.
\end{array} 
\]
\hfill $\Box$
\end{proof}
\section{Applications}\label{applications}
In this section we will describe how the general result of Theorem
\ref{charform} can be applied to concrete examples.

We will base these
results on variations of a labelled transition system defined as a
triple $\Psys=(\Proc,\Act, \der{})$, where
\begin{itemize}
\item $\Proc$ is a set,
\item $\Act$ is a set of labels and
\item $\der{}\subseteq \Proc\times\Act\times\Proc$ is a transition
relation. 
\end{itemize}
As usual, we  write $p\der{a}p'$ for $(p,a,p')\in \der{}$.
We often think of $\Proc$ as a set of processes, $\Act$ as a set of
actions, and $p\der{a}p'$ as a transition from process $p$ to process
$p'$ via action $a$.  We write $p \der{a}$ if there exists a $q$ such
that $p \der{a} q$, and we write $p\not\der{a}$ if there is no such
$q$.

All the characteristic-formula constructions we describe in this
section apply to labelled transition systems, and variations on that
model, for which the underlying set $\Proc$ and the set of labels $\Act$ are both finite.  

We will also use variations of the following language.  Given a set
$\Var$ of variables and a set $\Act$ of actions, we define the
language $\Lo(\Var,\Act)$ to be the standard Hennessy-Milner Logic
with recursion (HML)---see, for instance,~\cite{Larsen1990}---, given
by the grammar
\[
F::= \tr \mid \ff \mid X \mid F_1\land F_2 \mid F_1\lor F_2 \mid
\may{a}F_1 \mid \must{a} F_1,
\]
where $X\in \Var$ and $a\in \Act$.

Given a labelled transition system $\Psys = (\Proc,\Act,\der{})$, a
language $\Lo(\Var,\Act)$, and an environment $\sigma: \Var \to
\Po(\Proc)$, we define the semantics of the language by a relation
$\models$ between $\Env(\Var)\times \Proc$ and $\Lo(\Var,\Act)$, where
$(\sigma,p)\models F$ will be read as ``$F$ is true at $p$ with
respect to $\sigma$'', and we let $\not\models$ be the complement of
$\models$ in $(\Env(\Var)\times\Proc)\times\Lo(\Var,\Act)$.  The
relation $\models$ is defined by
\begin{center}
\begin{tabular}{l@{$\quad$iff$\quad$}l}
$(\sigma,p) \models \tr$ & $p\in \Proc$\\ $(\sigma,p) \models \ff$ &
$(\sigma,p) \not\models \tr$\\ $(\sigma,p) \models X$ & $p\in
\sigma(X)$\\ $(\sigma,p) \models F_1\land F_2$ & $(\sigma,p) \models
F_1$ and $(\sigma,p) \models F_2$\\ $(\sigma,p) \models F_1\lor F_2$ &
$(\sigma,p) \models F_1$ or $(\sigma,p) \models F_2$\\ $(\sigma,p)
\models \may{a}F_1$ & $(\sigma,p') \models F_1$ for some $p'$ for
which $p \der{a} p'$\\ $(\sigma,p) \models \must{a}F_1$ & $(\sigma,p')
\models F_1$ for all $p'$ for which $p \der{a} p'$.
\end{tabular}
\end{center}
One can easily check that the logic is monotonic and therefore the
largest and least fixed point constructions, as described in the
Section \ref{section:theorem},  naturally apply.
We use the standard abbreviations $\bigwedge_{i=1}^n F_i$ for $F_1\vee \cdots \vee F_n$ and $\bigvee_{i=1}^n F_i$ for $F_1\wedge \cdots \wedge F_n$.
We also set $\bigwedge_{i=1}^0 F_i \equiv \tr$ and $\bigvee_{i=1}^0F_i \equiv \ff$.
Because $\wedge$ and $\vee$ are commutative and associative, we may generally specify a finite index set rather than use an enumeration of formulas.
  
The following behavioural
equivalences are defined as the largest fixed points to monotonic functions.

\subsection{Simulation~\cite{Pa81}} 
Given $\Psys=(\Proc,\Act, \der{})$ and $S\subseteq\Proc \times \Proc$, let
$\F_{sim}(S)$ be defined such that 
\begin{quote}$(p,q)\in \F_{sim}(S)$ iff for
every $a\in \Act$ and $p'\in \Proc$,
\begin{itemize}
\item[] if $p\der{a} p'$ then there exists some $q'\in \Proc$ such that
$q\der{a}q'$ and $(p',q')\in S$.
\end{itemize}
\end{quote}
As $\F_{sim}$ is a monotonic function over $\Po(\Proc\times \Proc)$, by
Tarski's fixed point theorem, $\F$ has a largest fixed-point, which we
denote by $\sqlt_{sim}$.

We now search for characteristic formulas for $\sqlt_{sim}$ in the
language $\Lo(\Var,\Act)$, where $\Var = \{X_p\mid p\in \Proc\}$.  First
note that, for each $S\subseteq\Proc\times\Proc$,
\[
(p,q)\in\F_{sim}(S)\ioi (\sigma_S,p)\models \bigwedge_{a\in\Act}\must{a}
(\bigvee_{q'.q\der{a}q'}X_{q'}).
\]
Then, by Corollary \ref{cor:charform}, the characteristic formula for
$\sqlt_{sim}$ is given by the largest interpretation of the
declaration
\[
D_{sim}^\sqlt(X_q)=\bigwedge_{a\in\Act}\must{a}(\bigvee_{q'.q\der{a}q'}X_{q'}).
\]
The result above shows that to characterize $\sqlt_{sim}$ we only need
the fragment of HML that includes $\lor,\land,\must{a}$ for $a\in
\Act$, and a set of variables indexed over $\Proc$.

We now show how we can use Lemma \ref{inverse} to characterize
$\sqgt_{sim} = (\sqlt_{sim})^{-1}$, that is, where $p \sqgt_{sim} q$
if and only if $q \sqlt_{sim} p$.  The third component of the lemma
gives us $\sqgt_{sim} = \Fix(\F_{sim}^*)$.  We eventually aim to
characterize simulation equivalence, i.~e.~the intersection of two preorders,
and it is thus helpful to use a new set of variables $\Var' =
\{Y_q\mid q\in \Proc\}$ disjoint from $\Var$.  We get the following:
\[
\begin{array}{l}
(p,q)\in\F^*_{sim}(S)\ioi (q,p)\in\F_{sim}(S^{-1})\ioi\\\\
\forall a\in\Act,q'\in\Proc.\,q\der{a}q'\Rightarrow
\exists p'\in\Proc.p\der{a}p'\&(q',p')\in S^{-1}\ioi\\\\ 
\forall a\in\Act,q'\in\Proc.\,q\der{a}q'\Rightarrow
\exists p'\in\Proc.p\der{a}p'\&(p',q')\in S\ioi\\\\ 
(\sigma_S,p)\models \bigwedge_{a,q'.q\der{a}q'}\may{a}Y_{q'}.
\end{array}
\]
Then, by Corollary \ref{cor:charform}, the characteristic formula for
$\sqgt_{sim}$ is given as the largest interpretation of the
declaration
$$
D_{sim}^\sqgt(Y_q) = \bigwedge_{a,q'.q\der{a}q'}\may{a}Y_{q'}.
$$

To characterize $\sqgt_{sim}$ we only need the fragment of HML that
includes $\land,\may{a}$ for $a\in \Act$, and a set of variables
indexed over $\Proc$.

We can finally use these to define a characteristic formula for
simulation equivalence.  Define $\sim_{sim}$ such that $p\sim_{sim} q$
iff $p\sqlt_{sim} q$ and $p\sqgt_{sim} q$.  Then
$$ p \sim_{sim} q \ioi (\sigma^\sqlt_{\max}\uplus
\sigma^\sqgt_{\max},p) \models X_q \wedge Y_q,
$$ where
 $\sigma_1\uplus\sigma_2$ is defined on $domain(\sigma_1)\cup
domain(\sigma_2)$ if $domain(\sigma_1)\cap domain(\sigma_2)=\emptyset$
as
\[
\sigma_1\uplus\sigma_2(z)=\casedef{\sigma_1(z)}{z\in
domain(\sigma_1)}{\sigma_2(z)}.
\]
In this case we use the full logic HML.  

\subsection{Strong bisimulation~\cite{Pa81,Mi89}}
Given $\Psys=(\Proc,\Act, \der{})$ and $S\subseteq\Proc \times \Proc$, let
$\F_{bisim}(S)$ be defined such that
\begin{quote}
$(p,q)\in \F_{bisim}(S)$ iff for every $a\in \Act$, 
\begin{enumerate}
\item if $p\der{a}p'$, then there exists $q'\in\Proc$ such that
$q\der{a}q'$ and $(p',q')\in S$, and
\item if $q\der{a}q'$, then there exists $p'\in\Proc$ such that
$p\der{a}p'$ and $(p',q')\in S$.
\end{enumerate}
\end{quote}
As $\F_{bisim}$ is  monotonic, it has a largest  fixed point, which is
the  seminal notion  of  bisimulation equivalence  that  we denote  by
$\sim_{bisim}$.

As in the case of simulation, the first clause is translated into
\[
(\sigma_S,p) \models
\bigwedge_{a\in\Act}\must{a}(\bigvee_{q'.q\der{a}q'}X_{q'}),
\]
and the second one into 
\[
(\sigma_S,p)\models \bigwedge_{a,q'.q\der{a}q'}\may{a}X_{q'}.
\]
Then, by Corollary \ref{cor:charform}, the characteristic formula for
$\sim_{bisim}$ is given by the largest interpretation of the declaration.
\[
D_{bisim}(X_q) =
\bigwedge_{a\in\Act}\must{a}(\bigvee_{q'.q\der{a}q'}X_{q'})~\land
\bigwedge_{a,q'.q\der{a}q'}\may{a}X_{q'}.
\]
This is exactly the characteristic formula proposed in \cite{IGZ}.

\subsection{Ready simulation~\cite{BloomIM1995,LarsenS91}}
Given $\Psys=(\Proc,\Act, \der{})$ and $S\subseteq\Proc\times \Proc$, let
$\F_{RS}(S)$ be defined such that
\begin{quote}
$(p,q)\in \F_{RS}(S)$ iff for every $a\in \Act$ and $q'\in
\Proc$,
\begin{enumerate}
\item if $q\der{a}q'$, then there exists $p'\in \Proc$ such that
$p\der{a}p'$ and $(p',q')\in S$, and
\item if $p\der{a}$, then $q\der{a}$.
\end{enumerate}
\end{quote}
Clearly the second clause can be rewritten as ``if $q\not\der{a}$ then
$p\not\der{a}$''.  Also note that $\F_{RS}$ is monotonic.  We denote
the largest fixed point of $\F_{RS}$ by $\sqgt_{RS}$.\footnote{We
choose $\sqgt$ rather than the more familiar $\sqlt$ in order to both
use the commonly characteristic formula and establish greater
uniformity with the notation used elsewhere in the paper.}

In HML, $p\not\der{a}$ if and only if $p\models[a]\ff$.  Therefore,
for each $S\subseteq\Proc\times\Proc$, we have
\[
(p,q)\in\F_{RS}(S)\ioi (\sigma_S,p) \models
\bigwedge_{a,q'.q\der{a}q'}\may{a}X_{q'} ~\wedge
\bigwedge_{a.q\not\der{a}}\must{a}\ff.
\]
By Corollary \ref{cor:charform}, the characteristic formula for
$\sqgt_{RS}$ is now given as the largest interpretation of
$$
D_{RS}(X_q) = \bigwedge_{a,q'.q\der{a}q'}\may{a}X_{q'} ~\wedge
\bigwedge_{a.q\not\der{a}}\must{a}\ff.
$$

\subsection{Back and forth bisimulation}\label{section:backandforth}
So far our examples of applications of Theorem \ref{charform} have
only included known cases from the literature, i.~e.~where both the
semantic relation and its characteristic formula already exist. In
this section we will introduce a new semantic equivalence. This is a
variant of the back and forth bisimulation equivalence  introduced in
\cite{NMV90} that assumes several possible past states. 
The semantics introduced in \cite{NMV90} assumes that the past is
unique and consequently the derived equivalence coincides with the
standard strong bisimulation equivalence. This is not the case for the
multiple possible past semantics considered here. The introduction of
this behavioural equivalence serves as a stepping stone towards the one
introduced in the subsequent section.

Given $\Psys=(\Proc,\Act, \der{})$ and
$S\subseteq\Proc\times \Proc$, let $\F_{bfb}(S)$ be defined such that
\begin{quote}
$(p,q)\in \F_{bfb}(S)$ iff for
every $a\in \Act$
\begin{enumerate}\setlength{\itemsep}{0pt}
\item $\forall p'\in\Proc.\,p\der{a}p'\Rightarrow\exists
q'\in\Proc.q\der{a}q'\mbox{ and }(p',q')\in S$,
\item $\forall q'\in\Proc.\,q\der{a}q'\Rightarrow\exists
p'\in\Proc.p\der{a}p'\mbox{ and }(p',q')\in S$,
\item $\forall p'\in\Proc.\,p'\der{a}p\Rightarrow\exists
q'\in\Proc.q'\der{a}q\mbox{ and }(p',q')\in S$ and
\item $\forall q'\in\Proc.\,q'\der{a}q\Rightarrow\exists
p'\in\Proc.p'\der{a}p\mbox{ and }(p',q')\in S$.
\end{enumerate}
\end{quote}

To express such behaviour in the logical language considered so far,
we add two operators $\may{\overline{a}}$ and $\must{\overline{a}}$ to
it for every $a\in \Act$.  The semantics for these is given by
\begin{center} 
\begin{tabular}{l@{$\quad$iff$\quad$}l}
$(\sigma,p) \models \may{\overline{a}}F_1$ & $(\sigma,p')\models F_1$
for some $p'$ for which $p' \der{a} p$ and\\ $(\sigma,p) \models
\must{\overline{a}}F_1$ & $(\sigma,p')\models F_1$ for all $p'$ for
which $p' \der{a} p$.
\end{tabular}
\end{center}
Clearly these new operators are monotonic.  As in the case for
bisimulation equivalence, for each $S\subseteq\Proc\times\Proc$, the
first two clauses translate into
\[
(\sigma_S,p) \models
\bigwedge_{a\in\Act}\must{a}(\bigvee_{q'.q\der{a}q'}X_{q'})~\land
\bigwedge_{a,q'.q\der{a}q'}\may{a}X_{q'}.
\]
The second two clauses involve the new operators:
\[
(\sigma_S,p) \models
\bigwedge_{a\in\Act}\must{\overline{a}}(\bigvee_{q'.q'\der{a}q}X_{q'})~\land
\bigwedge_{a,q'.q'\der{a}q}\may{\overline{a}}X_{q'}.
\]
Then, by Corollary \ref{cor:charform}, the characteristic formula for
$\sim_{bfb}$ is given by the largest interpretation of the declaration
\begin{align*}
D_{bfb}(X_q) =\; &
\bigwedge_{a\in\Act}\must{a}(\bigvee_{q'.q\der{a}q'}X_{q'})\land
\bigwedge_{a,q'.q\der{a}q'}\may{a}X_{q'}\wedge \\ &
\bigwedge_{a\in\Act}\must{\overline{a}}(\bigvee_{q'.q'\der{a}q}X_{q'})\land
\bigwedge_{a,q'.q'\der{a}q}\may{\overline{a}}X_{q'}.
\end{align*}
\subsection{Back and forth bisimulation with indistinguishable states~\cite{MousaviLPAR07}}\label{section:backandforth-with-indistinguishable-states}
In this section we consider a version of the back and forth
bisimulation from the previous section where some of the states are
considered indistinguishable by some external agents.  For this
purpose we augment our notion of labelled transition systems with a
set $\Id$ of identities (or agents) and a family of equivalence
relation $\{\,\ind{i}\, \subseteq \Proc\times \Proc\mid
i\in\Id\}$. Such a structure is called an annotated labelled
transition system \cite{MousaviLPAR07} and is written as $(\Proc,\Act,
\der{A}, \Id, \ind{})$.

Given such a structure let $\F_{bfbid}(S)$ be defined such that
\begin{quote}
$(p,q)\in \F_{bfbid}(S)$ iff for every $a\in \Act$ and $i\in \Id$,
\begin{enumerate}\setlength{\itemsep}{0pt}
\item $\forall p'\in\Proc.\,p\der{a}p'\Rightarrow\exists
q'\in\Proc.q\der{a}q'\mbox{ and }(p',q')\in S$,
\item $\forall q'\in\Proc.\,q\der{a}q'\Rightarrow\exists
p'\in\Proc.p\der{a}p'\mbox{ and }(p',q')\in S$,
\item $\forall p'\in\Proc.\,p'\der{a}p\Rightarrow\exists
q'\in\Proc.q'\der{a}q\mbox{ and }(p',q')\in S$,
\item $\forall q'\in\Proc.\,q'\der{a}q\Rightarrow\exists
p'\in\Proc.p'\der{a}p\mbox{ and }(p',q')\in S$,
\item $\forall p'\in \Proc.\, p \ind{i} p'\Rightarrow \exists
q'\in\Proc.q\ind{i}q'\mbox{ and }(p',q')\in S$ and
\item $\forall q'\in\Proc.\,q\ind{i}q'\Rightarrow\exists
p'\in\Proc.p\ind{i}p'\mbox{ and }(p',q')\in S$.
\end{enumerate}
\end{quote}
We denote the largest fixed point of $\F_{bfbid}$ by $\sim_{bfbid}$.
We use the logical  language for back and forth bisimulation
from Section \ref{section:backandforth} and add to it the operators
$\may{i}$ and $\must{i}$ for each $i\in \Id$.  The semantics for these operators is given
by
\begin{center}
\begin{tabular}{l@{$\quad$iff$\quad$}l}
$(\sigma,p) \models \may{i}F_1$ & $(\sigma,p')\models F_1$ for some
$p'$ for which $p \ind{i} p'$ and\\ 
$(\sigma,p) \models \must{i}F_1$ &
$(\sigma,p')\models F_1$ for all $p'$ for which $p \ind{i} p'$.
\end{tabular}
\end{center}
These new constructions are clearly monotonic.  As for the case for
back and forth bisimulation, the first four clauses of $\F_{bfbid}(S)$
can be translated into
\begin{align*}
(\sigma_S,p) \models &
\bigwedge_{a\in\Act}\must{a}(\bigvee_{q'.q\der{a}q'}X_{q'})~\land
\bigwedge_{a,q'.q\der{a}q'}\may{a}X_{q'}~\wedge \\ &
\bigwedge_{a\in\Act}\must{\overline{a}}(\bigvee_{q'.q'\der{a}q}X_{q'})~\land
\bigwedge_{a,q'.q'\der{a}q}\may{\overline{a}}X_{q'}
\end{align*}
and the  last two clauses into
$$ (\sigma_S,p) \models
\bigwedge_{i\in\Id}\must{i}(\bigvee_{q'.q\ind{i}q'}X_{q'})~\land
\bigwedge_{i,q'.q\ind{i}q'}\may{i}X_{q'}.
$$ 
Then, by Corollary \ref{cor:charform}, the characteristic formula 
for $\sim_{bfbid}$ is given by the largest interpretation of the declaration
\begin{align*}
D_{bfb}(X_q) =\; &
\bigwedge_{a\in\Act}\must{a}(\bigvee_{q'.q\der{a}q'}X_{q'})~\land
\bigwedge_{a,q'.q\der{a}q'}\may{a}X_{q'}~\wedge \\ &
\bigwedge_{a\in\Act}\must{\overline{a}}(\bigvee_{q'.q'\der{a}q}X_{q'})~\land
\bigwedge_{a,q'.q'\der{a}q}\may{\overline{a}}X_{q'}~\wedge \\ &
\bigwedge_{i\in\Id}\must{i}(\bigvee_{q'.q\ind{i}q'}X_{q'})~\land
\bigwedge_{i,q'.q\ind{i}q'}\may{i}X_{q'}.
\end{align*}

As a consequence of the existence of this characteristic formula,
we obtain a behavioural characterization of the equivalence over
states in an annotated labelled transition system induced by the
epistemic logic studied in~\cite{MousaviLPAR07}. (More precisely, the
logic we study in this section may be seen as the positive version of
the one studied in~\cite{MousaviLPAR07}, where we use the modal
operator $\must{i}$ in lieu of $K_i$, read ``agent $i$ knows'', and
its dual.) This solves a problem that was left open in the
aforementioned reference.

\begin{theorem}
Let $p,q\in\Proc$. Then $p \mathbin{\sim_{bfbid}} q$ if, and only if, $p$ and $q$ satisfy the same formulae expressible in the logic considered in this section.
\end{theorem}

\subsection{Prebisimulation~\cite{Mi81,St87}}
We now extend the original labelled transition system with a
convergence predicate $\conv$ as found in \cite{Mi81}.  We write
$p\conv$ for $p\in \conv$, and we interpret $p\conv$ to mean that the
process $p$ converges.  If $p\not\in \conv$, we write $p\uparrow$, and
understand it to mean that the process $p$ diverges.

Given $(\Proc,\Act, \der{}, \conv)$ and $S\subseteq\Proc\times\Proc$, we define $\F_{prbis}(S)$ such that
\begin{quote}
$(p,q)\in \F_{prbis}(S)$ iff for every $a\in \Act$,
\begin{enumerate}
\item for all $q'\in\Proc$ if $q\der{a}q'$ then there exists
$p'\in\Proc$ such that $p\der{a}p'$ and $(p',q')\in S$, and
\item if $q\conv$, then both of the following hold:
\begin{enumerate}
\item $p\conv$ and
\item for all $p'\in \Proc$, if $p\der{a}p'$ then there exists $q'\in
\Proc$, such that $q\der{a}q'$ and $(p',q')\in S$.
\end{enumerate}
\end{enumerate}
\end{quote}
As $\F_{prbis}$ is monotonic, it has a largest fixed point
$\sqgt_{prbis}$,\footnote{For a similar reason as for the ready
simulation, we use $\sqgt_{prbis}$ rather than $\sqlt_{prbis}$.} known
as the prebisimulation preorder.  

To characterize this preorder we use an intuitionistic version
HML$_{int}$ of the standard HML. The syntax is the same as before, but
the definition for $\must{a}F$ is now
\begin{center}
\begin{tabular}{l@{$\quad$iff$\quad$}l}
$(\sigma,p) \models \must{a}F$ & $p\conv$ and $(\sigma,p')\models F$ for all $p'$ for which $p \der{a} p'$.
\end{tabular}
\end{center}

This implies that the first defining clause is the same as for
simulation
\[
(\sigma_S,p)\models \bigwedge_{a,q'.q\der{a}q'}\may{a}X_{q'},
\]
whereas the second one can be expressed as
\[
(\sigma_S,p)\models\casedef{\bigwedge_{a\in\Act}\must{a}
(\bigvee_{q'.q\der{a}q'}X_{q'})}{q\conv}{\tr}
\]
This can be  written differently as
\[
(\sigma_S,p)\models\bigwedge_{a\in\Act}\must{a}
(\bigvee_{q'.q\der{a}q'}X_{q'})
\lor\{\tr\mid q\divg\}, 
\]
where the notation $\lor\{\tr\mid q\divg\}$ means that the disjunct
$\tr$ is present only when $q\divg$.

Now, by Corollary \ref{cor:charform}, the characteristic formula for
$\sqgt_{prbis}$ is given as the largest interpretation of the declaration
\[
D_{prbis}(X_q) = \bigwedge_{a,q'.q\der{a}q'}\may{a}X_{q'} \wedge
[\bigwedge_{a\in\Act}\must{a}
(\bigvee_{q'.q\der{a}q'}X_{q'})
\lor\{\tr\mid q\divg\}].
\]

\subsection{Extended simulation~\cite{Thom87}}
We now consider labelled transition systems
augmented with a preorder relation $\sqlt_{\Act}$ over the set $\Act$
of labels.  Given $(\Proc,\Act, \der{}, \sqlt_{\Act})$ and
$S\subseteq\Proc\times \Proc$, we define $\F_{ext}$ such that
\begin{quote}
$(p,q)\in \F_{ext}(S)$ iff for every $a\in
\Act$,
\begin{itemize}
\item[] if $p\der{a} p'$ then there exists $q'\in \Proc$ and $b\in
\Act$ such that $a \sqlt_{\Act} b$, $q\der{b}q'$, and $(p',q')\in S$.
\end{itemize}
\end{quote}
We denote the largest fixed point of $\F_{ext}$ by $\sqlt_{ext}$.

To define the characteristic formula for $\sqlt_{ext}$ we use the
standard HML with recursion.  First note that for each
$S\subseteq\Proc\times\Proc$ and $p,q\in\Proc$
\[
(p,q)\in\F_{ext}(S)\ioi (\sigma_S,p)\models \bigwedge_{a\in\Act}\must{a}
(\bigvee_{b.a\sqlt_{\Act}b}\bigvee_{q'.q\der{b}q'}X_{q'}).
\]
By Corollary \ref{cor:charform}, the characteristic formula for
$\sqlt_{ext}$ is therefore given as the largest interpretation of
\[
D_{ext}^\sqlt(X_q)=\bigwedge_{a\in\Act}\must{a}(\bigvee_{b.a\sqlt_{\Act}b}\
\bigvee_{q'.q\der{b}q'}X_{q'}).
\]
As with simulation, we use Lemma \ref{inverse} to characterize
$\sqgt_{ext} = (\sqlt_{ext})^{-1}$.  We get that the characteristic
formula for this preorder is obtained as the largest interpretation of
the following declaration:
$$ D_{ext}^\sqgt(X_q) = \bigwedge_{a\in
\Act}\bigwedge_{q'.q\der{a}q'}\bigvee_{b.a\sqlt_{\Act}b}\may{b}X_{q'}.
$$

\section{Conclusion}\label{conclusion}
This paper provides a general view of characteristic formulae for
suitable behavioural relations.  The relations of interest are those
that can be defined by largest or smallest fixed points of monotonic
functions, which can be expressed by declarations over a given
language.  Theorem \ref{charform} shows that a declaration that
expresses such a function can be viewed as the characteristic formula
of either the largest or least fixed point of the function.  We have
explored a number of applications of this theorem, some in recovering
characteristic formulae already discovered, and some being novel
constructions.  But each of the behavioural relations we consider are
largest fixed points of functions, and we hope future work can yield
characteristic formulae for interesting least fixed points as well.
There are still, however, many other largest fixed points that may be
applications of this theorem.  These include weak bisimulation
equivalence~\cite{Mi89}, weak bisimulation congruence~\cite{Mi89},
branching bisimulation equivalence~\cite{JACM::GlabbeekW1996},
resource bisimulation equivalence~\cite{CorradiniNL99}, and
g-bisimulation equivalence~\cite{rijk:note00}. 
\bibliographystyle{eptcs}
\bibliography{abbreviations,char-general,proceedings}

\begin{thebibliography}{10}
\providecommand{\bibitemstart}[1]{\bibitem{#1}}
\providecommand{\bibitemend}{}
\providecommand{\bibliographystart}{}
\providecommand{\bibliographyend}{}
\providecommand{\url}[1]{\texttt{#1}}
\providecommand{\urlprefix}{Available at }
\providecommand{\bibinfo}[2]{#2}
\bibliographystart

\bibitemstart{AILS2007}
\bibinfo{author}{Luca Aceto}, \bibinfo{author}{Anna Ingolfsdottir},
  \bibinfo{author}{Kim~G. Larsen} \& \bibinfo{author}{Ji\v{r}\'{\i} Srba}
  (\bibinfo{year}{2007}): \emph{\bibinfo{title}{Reactive Systems: Modelling,
  Specification and Verification}}.
\newblock \bibinfo{publisher}{Cambridge University Press}.
\bibitemend

\bibitemstart{AcetoIPP00}
\bibinfo{author}{Luca Aceto}, \bibinfo{author}{Anna Ing{o}lfsd{o}ttir},
  \bibinfo{author}{Mikkel~Lykke Pedersen} \& \bibinfo{author}{Jan Poulsen}
  (\bibinfo{year}{2000}): \emph{\bibinfo{title}{Characteristic formulae for
  timed automata}}.
\newblock {\sl \bibinfo{journal}{RAIRO, Theoretical Informatics and
  Applications}} \bibinfo{volume}{34}(\bibinfo{number}{6}), pp.
  \bibinfo{pages}{565--584}.
\newblock \urlprefix\url{http://dx.doi.org/10.1051/ita:2000131}.
\bibitemend

\bibitemstart{BloomIM1995}
\bibinfo{author}{Bard Bloom}, \bibinfo{author}{Sorin Istrail} \&
  \bibinfo{author}{Albert Meyer} (\bibinfo{year}{1995}):
  \emph{\bibinfo{title}{Bisimulation can't be Traced}}.
\newblock {\sl \bibinfo{journal}{Journal of the ACM}}
  \bibinfo{volume}{42}(\bibinfo{number}{1}), pp. \bibinfo{pages}{232--268}.
\bibitemend

\bibitemstart{BoudolL1992}
\bibinfo{author}{G{\'e}rard Boudol} \& \bibinfo{author}{Kim~G. Larsen}
  (\bibinfo{year}{1992}): \emph{\bibinfo{title}{Graphical versus logical
  specifications}}.
\newblock {\sl \bibinfo{journal}{Theoretical Computer Science}}
  \bibinfo{volume}{106}(\bibinfo{number}{1}), pp. \bibinfo{pages}{3--20}.
\bibitemend

\bibitemstart{BCG88}
\bibinfo{author}{M.C. Browne}, \bibinfo{author}{E.M. Clarke} \&
  \bibinfo{author}{O.~Gr\"{u}mberg} (\bibinfo{year}{1988}):
  \emph{\bibinfo{title}{Characterizing finite {Kripke} structures in
  propositional temporal logic}}.
\newblock {\sl \bibinfo{journal}{Theoretical Computer Science}}
  \bibinfo{volume}{59}(\bibinfo{number}{1,2}), pp. \bibinfo{pages}{115--131}.
\bibitemend

\bibitemstart{ClarkeGP1999}
\bibinfo{author}{Ed~Clarke}, \bibinfo{author}{Orna Gruemberg} \&
  \bibinfo{author}{Doron Peled} (\bibinfo{year}{1999}):
  \emph{\bibinfo{title}{Model Checking}}.
\newblock \bibinfo{publisher}{MIT Press}.
\bibitemend

\bibitemstart{ClEm81}
\bibinfo{author}{E.M. Clarke} \& \bibinfo{author}{E.A. Emerson}
  (\bibinfo{year}{1981}): \emph{\bibinfo{title}{Design and Synthesis of
  Synchronization Skeletons using Branching Time Temporal Logic}}.
\newblock In: \bibinfo{editor}{{D. Kozen}}, editor: {\sl
  \bibinfo{booktitle}{Proceedings of the Workshop on Logics of Programs}}, {\sl
  \bibinfo{series}{Lecture Notes in Computer Science}} \bibinfo{volume}{131}.
  \bibinfo{publisher}{Springer-Verlag}, pp. \bibinfo{pages}{52--71}.
\bibitemend

\bibitemstart{CleavelandS1991}
\bibinfo{author}{Rance Cleaveland} \& \bibinfo{author}{Bernhard Steffen}
  (\bibinfo{year}{1991}): \emph{\bibinfo{title}{Computing Behavioural
  Relations, Logically}}.
\newblock In: \bibinfo{editor}{J.~Leach~Albert}, \bibinfo{editor}{B.~Monien} \&
  \bibinfo{editor}{M.~Rodr\'{\i}guez}, editors: {\sl
  \bibinfo{booktitle}{Proceedings $18^{th}$ ICALP, {\rm Madrid}}}, {\sl
  \bibinfo{series}{Lecture Notes in Computer Science}} \bibinfo{volume}{510}.
  \bibinfo{publisher}{Springer-Verlag}, pp. \bibinfo{pages}{127--138}.
\bibitemend

\bibitemstart{CorradiniNL99}
\bibinfo{author}{Flavio Corradini}, \bibinfo{author}{Rocco~De Nicola} \&
  \bibinfo{author}{Anna Labella} (\bibinfo{year}{1999}):
  \emph{\bibinfo{title}{Graded Modalities and Resource Bisimulation}}.
\newblock In: \bibinfo{editor}{C.~Pandu Rangan}, \bibinfo{editor}{Venkatesh
  Raman} \& \bibinfo{editor}{Ramaswamy Ramanujam}, editors: {\sl
  \bibinfo{booktitle}{Foundations of Software Technology and Theoretical
  Computer Science, 19th Conference, Chennai, India, December 13-15, 1999,
  Proceedings}}, {\sl \bibinfo{series}{Lecture Notes in Computer Science}}
  \bibinfo{volume}{1738}. \bibinfo{publisher}{Springer-Verlag}, pp.
  \bibinfo{pages}{381--393}.
\newblock
  \urlprefix\url{http://link.springer.de/link/service/series/0558/bibs/1738/17%
380381.htm}.
\bibitemend

\bibitemstart{NMV90}
\bibinfo{author}{Rocco De~Nicola}, \bibinfo{author}{Ugo Montanari} \&
  \bibinfo{author}{Frits Vaandrager} (\bibinfo{year}{1990}):
  \emph{\bibinfo{title}{Back and forth bisimulations}}.
\newblock In: {\sl \bibinfo{booktitle}{C{ONCUR}' 90 ({A}msterdam, 1990)}}, {\sl
  \bibinfo{series}{Lecture Notes in Comput. Sci.}} \bibinfo{volume}{458}.
  \bibinfo{publisher}{Springer}, \bibinfo{address}{Berlin}, pp.
  \bibinfo{pages}{152--165}.
\bibitemend

\bibitemstart{MousaviLPAR07}
\bibinfo{author}{Francien Dechesne}, \bibinfo{author}{MohammadReza Mousavi} \&
  \bibinfo{author}{Simona Orzan} (\bibinfo{year}{2007}):
  \emph{\bibinfo{title}{Operational and Epistemic Approaches to Protocol
  Anlaysis: Bridging the Gap}}.
\newblock In: {\sl \bibinfo{booktitle}{Proceedings of the 14th International
  Conference on Logic for Programming Artificial Intelligence and Reasoning
  {(LPAR'07)}}}, {\sl \bibinfo{series}{Lecture Notes in Artificial
  Intelligence}} \bibinfo{volume}{4790}. \bibinfo{publisher}{Springer-Verlag},
  pp. \bibinfo{pages}{226--241}.
\bibitemend

\bibitemstart{Emerson1990}
\bibinfo{author}{E.~Allen Emerson} (\bibinfo{year}{1990}):
  \emph{\bibinfo{title}{Temporal and modal logic}}.
\newblock In: {\sl \bibinfo{booktitle}{Handbook of Theoretical Computer
  Science, Vol.\ B}}. \bibinfo{publisher}{Elsevier},
  \bibinfo{address}{Amsterdam}, pp. \bibinfo{pages}{995--1072}.
\bibitemend

\bibitemstart{FecherS:bib:05:CFUTS}
\bibinfo{author}{Harald Fecher} \& \bibinfo{author}{Martin Steffen}
  (\bibinfo{year}{2005}): \emph{\bibinfo{title}{Characteristic $\mu$-Calculus
  Formula for an Underspecified Transition System}}.
\newblock In: {\sl \bibinfo{booktitle}{EXPRESS'04}}, {\sl
  \bibinfo{series}{Electronic Notes in Theoretical Computer Science}}
  \bibinfo{volume}{128}. \bibinfo{publisher}{Elsevier Science Publishers}, pp.
  \bibinfo{pages}{103--116}.
\newblock
  \urlprefix\url{http://www.informatik.uni-kiel.de/~hf/papers/Fecher04express.%
pdf}.
\bibitemend

\bibitemstart{vG2001}
\bibinfo{author}{R.~{van} Glabbeek} (\bibinfo{year}{2001}):
  \emph{\bibinfo{title}{The linear time--branching time spectrum. {I}. {T}he
  semantics of concrete, sequential processes}}.
\newblock In: \bibinfo{editor}{Jan Bergstra}, \bibinfo{editor}{Alban Ponse} \&
  \bibinfo{editor}{Scott~A. Smolka}, editors: {\sl \bibinfo{booktitle}{Handbook
  of Process Algebra}}. \bibinfo{publisher}{Elsevier}, pp.
  \bibinfo{pages}{3--99}.
\bibitemend

\bibitemstart{JACM::GlabbeekW1996}
\bibinfo{author}{R.~{van} Glabbeek} \& \bibinfo{author}{W.P. Weijland}
  (\bibinfo{year}{1996}): \emph{\bibinfo{title}{Branching Time and Abstraction
  in Bisimulation Semantics}}.
\newblock {\sl \bibinfo{journal}{Journal of the ACM}}
  \bibinfo{volume}{43}(\bibinfo{number}{3}), pp. \bibinfo{pages}{555--600}.
\bibitemend

\bibitemstart{GrafS86a}
\bibinfo{author}{S.~Graf} \& \bibinfo{author}{J.~Sifakis}
  (\bibinfo{year}{1986}): \emph{\bibinfo{title}{A Modal Characterization of
  Observational Congruence on Finite Terms of {CCS}}}.
\newblock {\sl \bibinfo{journal}{Information and Control}}
  \bibinfo{volume}{68}(\bibinfo{number}{1--3}), pp. \bibinfo{pages}{125--145}.
\bibitemend

\bibitemstart{HM85}
\bibinfo{author}{M.~Hennessy} \& \bibinfo{author}{R.~Milner}
  (\bibinfo{year}{1985}): \emph{\bibinfo{title}{Algebraic laws for
  nondeterminism and concurrency}}.
\newblock {\sl \bibinfo{journal}{Journal of the ACM}}
  \bibinfo{volume}{32}(\bibinfo{number}{1}), pp. \bibinfo{pages}{137--161}.
\bibitemend

\bibitemstart{IGZ}
\bibinfo{author}{Anna Ingolfsdottir}, \bibinfo{author}{Jens~Christian
  Godskesen} \& \bibinfo{author}{Michael Zeeberg} (\bibinfo{year}{1987}):
  \emph{\bibinfo{title}{Fra {Hennessy-Milner} Logik til {CCS}-Processer}}.
\newblock \bibinfo{type}{Master's thesis}, \bibinfo{school}{Department of
  Computer Science, Aalborg University}.
\newblock \bibinfo{note}{In Danish}.
\bibitemend

\bibitemstart{Ke76}
\bibinfo{author}{R.M. Keller} (\bibinfo{year}{1976}):
  \emph{\bibinfo{title}{Formal verification of parallel programs}}.
\newblock {\sl \bibinfo{journal}{Communications of the ACM}}
  \bibinfo{volume}{19}(\bibinfo{number}{7}), pp. \bibinfo{pages}{371--384}.
\bibitemend

\bibitemstart{Kor91}
\bibinfo{author}{H.~Korver} (\bibinfo{year}{1992}):
  \emph{\bibinfo{title}{Computing Distinguishing Formulas for Branching
  Bisimulation}}.
\newblock In: \bibinfo{editor}{K.G. Larsen} \& \bibinfo{editor}{A.~Skou},
  editors: {\sl \bibinfo{booktitle}{Proceedings of the Third Workshop on
  Computer Aided Verification, {\rm Aalborg, Denmark, July 1991}}}, {\sl
  \bibinfo{series}{Lecture Notes in Computer Science}} \bibinfo{volume}{575}.
  \bibinfo{publisher}{Springer-Verlag}, pp. \bibinfo{pages}{13--23}.
\bibitemend

\bibitemstart{llw95}
\bibinfo{author}{F.~Laroussinie}, \bibinfo{author}{K.~G. Larsen} \&
  \bibinfo{author}{C.~Weise} (\bibinfo{year}{1995}): \emph{\bibinfo{title}{From
  Timed Automata to Logic - and Back}}.
\newblock In: \bibinfo{editor}{Jir{\'\i} Wiedermann} \& \bibinfo{editor}{Petr
  H{\'a}jek}, editors: {\sl \bibinfo{booktitle}{Mathematical Foundations of
  Computer Science 1995, 20th International Symposium}}, {\sl
  \bibinfo{series}{Lecture Notes in Computer Science}} \bibinfo{volume}{969}.
  \bibinfo{publisher}{Springer}, \bibinfo{address}{Prague, Czech Republic}, pp.
  \bibinfo{pages}{529--539}.
\bibitemend

\bibitemstart{Larsen1990}
\bibinfo{author}{Kim~Guldstrand Larsen} (\bibinfo{year}{1990}):
  \emph{\bibinfo{title}{Proof Systems for Satisfiability in {Hennessy--Milner}
  Logic with Recursion}}.
\newblock {\sl \bibinfo{journal}{Theoretical Computer Science}}
  \bibinfo{volume}{72}(\bibinfo{number}{2--3}), pp. \bibinfo{pages}{265--288}.
\bibitemend

\bibitemstart{LS92}
\bibinfo{author}{Kim~Guldstrand Larsen} \& \bibinfo{author}{A.~Skou}
  (\bibinfo{year}{1992}): \emph{\bibinfo{title}{Compositional Verification of
  Probabilistic Processes}}.
\newblock In: \bibinfo{editor}{Rance Cleaveland}, editor: {\sl
  \bibinfo{booktitle}{Proceedings CONCUR 92, {\rm Stony Brook, NY, USA}}}, {\sl
  \bibinfo{series}{Lecture Notes in Computer Science}} \bibinfo{volume}{630}.
  \bibinfo{publisher}{Springer-Verlag}, pp. \bibinfo{pages}{456--471}.
\bibitemend

\bibitemstart{LarsenS91}
\bibinfo{author}{Kim~Gulstrand Larsen} \& \bibinfo{author}{A.~Skou}
  (\bibinfo{year}{1991}): \emph{\bibinfo{title}{Bisimulation through
  Probabilistic Testing}}.
\newblock {\sl \bibinfo{journal}{Information and Computation}}
  \bibinfo{volume}{94}(\bibinfo{number}{1}), pp. \bibinfo{pages}{1--28}.
\bibitemend

\bibitemstart{MaSt93}
\bibinfo{author}{Tiziana Margaria} \& \bibinfo{author}{Bernhard Steffen}
  (\bibinfo{year}{1993}): \emph{\bibinfo{title}{Distinguishing Formulas for
  Free}}.
\newblock In: {\sl \bibinfo{booktitle}{Proc. EDAC--EUROASIC'93: IEEE European
  Design Automation Conference, Paris (France)}}. \bibinfo{publisher}{IEEE
  Computer Society Press}.
\bibitemend

\bibitemstart{Mi81}
\bibinfo{author}{R.~Milner} (\bibinfo{year}{1981}): \emph{\bibinfo{title}{A
  modal characterisation of observable machine behaviour}}.
\newblock In: \bibinfo{editor}{E.~Astesiano} \& \bibinfo{editor}{C.~B{\"o}hm},
  editors: {\sl \bibinfo{booktitle}{CAAP '81: Trees in Algebra and Programming,
  6th Colloquium}}, {\sl \bibinfo{series}{Lecture Notes in Computer Science}}
  \bibinfo{volume}{112}. \bibinfo{publisher}{Springer-Verlag}, pp.
  \bibinfo{pages}{25--34}.
\bibitemend

\bibitemstart{Mi89}
\bibinfo{author}{R.~Milner} (\bibinfo{year}{1989}):
  \emph{\bibinfo{title}{Communication and Concurrency}}.
\newblock \bibinfo{publisher}{Prentice-Hall International},
  \bibinfo{address}{Englewood Cliffs}.
\bibitemend

\bibitemstart{MullerOlm1998}
\bibinfo{author}{Markus M{\"u}ller-Olm} (\bibinfo{year}{1998}):
  \emph{\bibinfo{title}{Derivation of Characteristic Formulae}}.
\newblock In: {\sl \bibinfo{booktitle}{MFCS'98 Workshop on Concurrency (Brno,
  1998)}}, {\sl \bibinfo{series}{Electron. Notes Theor. Comput.
  Sci.}}~\bibinfo{volume}{18}. \bibinfo{publisher}{Elsevier},
  \bibinfo{address}{Amsterdam}, p. \bibinfo{pages}{12 pp. (electronic)}.
\bibitemend

\bibitemstart{Pa81}
\bibinfo{author}{D.~Park} (\bibinfo{year}{1981}):
  \emph{\bibinfo{title}{Concurrency and automata on infinite sequences}}.
\newblock In: \bibinfo{editor}{P.~Deussen}, editor: {\sl
  \bibinfo{booktitle}{5th GI Conference, {\rm Karlsruhe, Germany}}}, {\sl
  \bibinfo{series}{Lecture Notes in Computer Science}} \bibinfo{volume}{104}.
  \bibinfo{publisher}{Springer-Verlag}, pp. \bibinfo{pages}{167--183}.
\bibitemend

\bibitemstart{Plotkin04a}
\bibinfo{author}{Gordon~D. Plotkin} (\bibinfo{year}{2004}):
  \emph{\bibinfo{title}{A Structural Approach to Operational Semantics}}.
\newblock {\sl \bibinfo{journal}{Journal of Logic and Algebraic Programming}}
  \bibinfo{volume}{60--61}, pp. \bibinfo{pages}{17--139}.
\bibitemend

\bibitemstart{Pnueli77}
\bibinfo{author}{Amir Pnueli} (\bibinfo{year}{1977}): \emph{\bibinfo{title}{The
  Temporal Logic of Programs}}.
\newblock In: {\sl \bibinfo{booktitle}{Proceedings $\it 18^{th}$ Annual
  Symposium on Foundations of Computer Science}}. \bibinfo{publisher}{IEEE},
  pp. \bibinfo{pages}{46--57}.
\bibitemend

\bibitemstart{QS81}
\bibinfo{author}{J.~P. Queille} \& \bibinfo{author}{J.~Sifakis}
  (\bibinfo{year}{1981}): \emph{\bibinfo{title}{Specification and Verification
  of Concurrent Systems in {Cesar}}}.
\newblock In: {\sl \bibinfo{booktitle}{Proceedings of the 5th International
  Symposium on Programming}}, {\sl \bibinfo{series}{Lecture Notes in Computer
  Science}} \bibinfo{volume}{137}. \bibinfo{publisher}{Springer-Verlag}, pp.
  \bibinfo{pages}{337--351}.
\bibitemend

\bibitemstart{rijk:note00}
\bibinfo{author}{M.~de~Rijke} (\bibinfo{year}{2000}): \emph{\bibinfo{title}{A
  Note on Graded Modal Logic}}.
\newblock {\sl \bibinfo{journal}{Studia Logica}}
  \bibinfo{volume}{64}(\bibinfo{number}{2}), pp. \bibinfo{pages}{271--283}.
\bibitemend

\bibitemstart{SteffenI94}
\bibinfo{author}{Bernhard Steffen} \& \bibinfo{author}{Anna Ing{o}lfsd{o}ttir}
  (\bibinfo{year}{1994}): \emph{\bibinfo{title}{Characteristic Formulae for
  Processes with Divergence}}.
\newblock {\sl \bibinfo{journal}{Information and Computation}}
  \bibinfo{volume}{110}(\bibinfo{number}{1}), pp. \bibinfo{pages}{149--163}.
\bibitemend

\bibitemstart{St87}
\bibinfo{author}{Colin Stirling} (\bibinfo{year}{1987}):
  \emph{\bibinfo{title}{Modal logics for communicating systems}}.
\newblock {\sl \bibinfo{journal}{Theoret. Comput. Sci.}}
  \bibinfo{volume}{49}(\bibinfo{number}{2-3}), pp. \bibinfo{pages}{311--347}.
\newblock \bibinfo{note}{Twelfth international colloquium on automata,
  languages and programming (Nafplion, 1985)}.
\bibitemend

\bibitemstart{Ta55}
\bibinfo{author}{A.~Tarski} (\bibinfo{year}{1955}): \emph{\bibinfo{title}{A
  Lattice-Theoretical Fixpoint Theorem and its Applications}}.
\newblock {\sl \bibinfo{journal}{Pacific Journal of Mathematics}}
  \bibinfo{volume}{5}, pp. \bibinfo{pages}{285--309}.
\bibitemend

\bibitemstart{Thom87}
\bibinfo{author}{B.~Thomsen} (\bibinfo{year}{1987}): \emph{\bibinfo{title}{An
  extended bisimulation induced by a preorder on actions}}.
\newblock \bibinfo{type}{Master's thesis}, \bibinfo{school}{Aalborg
  {University} {Centre}}.
\bibitemend

\bibliographyend
\end{thebibliography}

\end{document}